\documentclass[aps, reprint,superscriptaddress, prl, noeprint]{revtex4-2}

\usepackage{graphicx}
\usepackage{dcolumn} 
\usepackage{bm}
\usepackage{amssymb}
\usepackage{amsmath}
\usepackage{color}

\usepackage{xr}
\newcommand{\FigCap}[1]{(#1)}	

\newcommand{\ts}{$T_{{\rm s}}$}
\newcommand{\tc}{$T_{{\rm c}}$}
\newcommand{\hc}{$H_{{\rm c2}}$}
\newcommand{\fesete}{FeSe$_{1-x}$Te$_{x}$}
\newcommand{\feses}{FeSe$_{1-x}$S$_{x}$}

\begin{document}

\title{Enhanced superconducting pairing strength near a nonmagnetic nematic quantum critical point}

\author{K.~Mukasa}
	\thanks{These authors have contributed equally to this work.}
\author{K.~Ishida}
	\thanks{These authors have contributed equally to this work.}
	\altaffiliation[Present address: ]{Max Planck Institute for Chemical Physics of Solids, N\"{o}thnitzer Stra{\ss}e 40, 01187 Dresden, Germany}
	\email{kousuke.ishida@cpfs.mpg.de}
	\affiliation{Department of Advanced Materials Science, University of Tokyo, Kashiwa, Chiba 277-8561, Japan}
\author{S.~Imajo}
    \affiliation{Institute for Solid State Physics, University of Tokyo, Kashiwa, Chiba, 277-8581, Japan}
\author{M.~W.~Qiu}
\author{M.~Saito}
\author{K.~Matsuura}
    \altaffiliation[Present address: ]{Research Center for Advanced Science and Technology (RCAST), University of Tokyo, Meguro-ku, Tokyo 153-8904, Japan}
\author{Y.~Sugimura}
\author{S.~Liu}
    \affiliation{Department of Advanced Materials Science, University of Tokyo, Kashiwa, Chiba 277-8561, Japan}
\author{Y.~Uezono}
\author{T.~Otsuka}
	\affiliation{Graduate School of Science and Technology, Hirosaki University, Hirosaki, Aomori 036-8561, Japan}
\author{M.~\v{C}ulo}
	\altaffiliation[Permanent address: ]{Institut za ﬁziku, P.O. Box 304, HR-10001 Zagreb, Croatia}
	\affiliation{High Field Magnet Laboratory (HFML-EMFL) and Institute for Molecules and Materials, Radboud University, Toernooiveld 7, 6525 ED Nijmegen, Netherlands}
\author{S.~Kasahara}
    \affiliation{Department of Physics, Kyoto University, Kyoto 606-8502, Japan}
    \affiliation{Research Institute for Interdisciplinary Science, Okayama University, Okayama 700-8530, Japan}
\author{Y.~Matsuda}
	\affiliation{Department of Physics, Kyoto University, Kyoto 606-8502, Japan}
\author{N.~E.~Hussey}
	\affiliation{High Field Magnet Laboratory (HFML-EMFL) and Institute for Molecules and Materials, Radboud University, Toernooiveld 7, 6525 ED Nijmegen, Netherlands}
	\affiliation{H. H. Wills Physics Laboratory, University of Bristol, Tyndall Avenue, Bristol BS8 1TL, United Kingdom}
\author{T.~Watanabe}
    \affiliation{Graduate School of Science and Technology, Hirosaki University, Hirosaki, Aomori 036-8561, Japan}
\author{K.~Kindo}
    \affiliation{Institute for Solid State Physics, University of Tokyo, Kashiwa, Chiba, 277-8581, Japan}
\author{T.~Shibauchi}
    \email{shibauchi@k.u-tokyo.ac.jp}
    \affiliation{Department of Advanced Materials Science, University of Tokyo, Kashiwa, Chiba 277-8561, Japan}

\date{\today}


\begin{abstract}
The quest for high-temperature superconductivity at ambient pressure is a central issue in physics. 
In this regard, the relationship between unconventional superconductivity and the quantum critical point (QCP) associated with the suppression of some form of symmetry-breaking order to zero temperature has received particular attention. 
The key question is how the strength of the electron pairs changes near the QCP, and this can be verified by high-field experiments. 
However, such studies are limited mainly to superconductors with magnetic QCPs, and the possibility of unconventional mechanisms by which nonmagnetic QCP promotes strong pairing remains a nontrivial issue. 
Here, we report systematic measurements of the upper critical field \hc\, in nonmagnetic \fesete\, superconductors, which exhibit a QCP of electronic nematicity characterized by spontaneous rotational-symmetry breaking. 
As the magnetic field increases, the superconducting phase of \fesete\, shrinks to a narrower dome surrounding the nematic QCP. 
The analysis of \hc\, reveals that the Pauli-limiting field is enhanced toward the QCP, implying that the pairing interaction is significantly strengthened via nematic fluctuations emanated from the QCP. 
Remarkably, this nonmagnetic nematic QCP is not accompanied by a divergent effective mass, distinct from the magnetically mediated pairing. 
Our observation opens up a nonmagnetic route to high-temperature superconductivity.

\end{abstract}

\maketitle
 

	Unconventional superconductors, whose mechanisms of electron pairing are distinct from the conventional electron-phonon interaction, have been a focus of interest in condensed matter physics.
	It has been increasingly recognized that most of their parent materials undergo a phase transition toward a long-range electronic order.
	The superconducting transition temperature \tc\, often reaches a maximum at a quantum critical point (QCP), where a continuous order-disorder transition of coexisting order takes place at absolute zero temperature. 
	This implies that quantum fluctuations intensified at the QCP might strengthen the interaction binding Cooper pairs, leading to the enlarged magnitude of superconducting gap at the ground state.
	Mapping out the upper critical field \hc, especially the Pauli-limiting field which is a fundamental measure of superconducting condensation energy \cite{Clogston1962,Chandrasekhar1962}, across the superconducting phase diagram is a direct test for the impact of QCP on superconductivity.
	However, in many cases, \hc\, is governed by the orbital pair breaking effect, and it often requires a very high magnetic field, which makes such evaluations difficult. 

	While an increase of \tc\, has been often observed near the QCP of antiferromagnetic orders \cite{Shibauchi2014,Mathur1998}, electronic nematicity, which breaks a rotational symmetry of the underlying lattice with preserving a translational symmetry, has been found in several classes of unconventional superconductors including high-\tc\, copper-oxides \cite{Sato2017, Ishida2020}, heavy-fermion compounds \cite{Okazaki2011, Ronning2017}, and iron-based superconductors \cite{Fernandes2014}.
	Although it has been theoretically proposed that nematic fluctuations can have a cooperative relation to superconductivity \cite{Kontani2010, Lederer2015, Lederer2017}, the enhancement of \tc\, associated with the nematic QCP in these compounds has been still under debate.
	This is mostly because the electronic nematic orders found in the correlated systems often intertwine with other charge or spin degrees of freedom, and they do not exist in isolation. 

	In this context, the iron-chalcogenide superconductor FeSe is an important material \cite{Shibauchi2020, Kreisel2020, Coldea2021}.
	Below $T_{\rm s} \approx 90$\,K, FeSe exhibits a unique nematic order, which does not accompany a long-range magnetic order, unlike other iron-based superconductors \cite{Bohmer2013}.
	This pure nematic order can be controlled by the physical pressure \cite{Sun2016} and the isovalent substitution of S for Se sites \cite{Hosoi2016}. 
	In both cases, however, there is no evidence for the \tc\, enhancement associated with the QCP  of the nematic phase.
	In the temperature versus pressure phase diagrams of \feses\, \cite{Matsuura2017}, \tc\, is less affected by the nematic endpoint, while \tc\, increases above 30\,K at the point where the pressure-induced magnetic order disappears.
	At ambient pressure, $T_{\rm c}$ of FeSe$_{1-x}$S$_x$ shows a broad maximum at $x \approx 0.10$ inside the nematic phase but decreases to $\approx 4$\,K above the nematic QCP at $x \approx 0.17$ \cite{Mizukami2021}. 
	Nuclear magnetic resonance (NMR) measurements reveal a correlation between $T_{\rm c}$ and the strength of spin-lattice relaxation rate $1/T_1T$ inside the nematic phase, suggesting the importance of spin fluctuations despite the absence of long-range magnetic order in FeSe$_{1-x}$S$_x$ \cite{Wiecki2018}. 

	Recently, in contrast to the S substitution and physical pressure cases, \fesete\,has been found to show an enhancement of \tc\, near the endpoint of the nematic phase at $x \approx 0.50$ without static magnetism \cite{Mukasa2021}.
	With Te substitution, \tc\, first decreases to its local minimum at $x \approx 0.30$ and then turns to increase toward a broad maximum of at $x \approx $ 0.60 with $T_{\rm c} \approx $ 14\,K.
	Elastoresistivity studies have shown that the nematic susceptibility of \fesete\, follows a Curie-Weiss temperature dependence over a wide region of the phase diagram, evidencing the presence of a nonmagnetic nematic QCP \cite{Ishida2022}.
	Moreover, the superconducting dome of \fesete\, straddles the nematic QCP, making it the first viable example of a system exhibiting a link between pure nematic fluctuations and enhanced superconductivity.
	To verify the notion of electron pairing promoted by nematic fluctuations, however, it is essential to clarify how the pairing strength evolves in the phase diagram. 
	In this study, we present systematic studies of superconductivity in FeSe-based materials at high magnetic fields up to 60\,T.
	Thanks to their modest \tc, we have succeeded in completely suppressing the superconducting phase across the entire phase diagram.
	Single crystals of \feses\, and \fesete\, for $0 < x \leq 0.48$ were grown by the chemical vapor transport (CVT) technique \cite{Mukasa2021}.
	Single crystals of \fesete\, for $0.52 \leq x \leq 0.90$ were obtained by the Bridgman method \cite{Watanabe2020}.
	For the  crystals synthesized by the Bridgman method, the Te annealing procedure was applied to minimize the excess Fe \cite{Watanabe2020}.
	The actual Te composition $x$ of \fesete\, synthesized by the CVT method was determined for each sample from the $c$-axis length measured by X-ray diffraction measurements. 
	The $x$ values of crystals grown by the Bridgman method are taken from the nominal values.
    
    High-field electrical resistance and tunnel diode oscillator (TDO) measurements of \fesete\, were carried out at International MegaGauss Science Laboratory, Institute for Solid State Physics in University of Tokyo, using a 60-T pulsed magnet.
	In the TDO measurements, a TDO circuit with a 0.7 mm-diameter 8-shaped copper coil ($\approx$ 8 turns to cancel out the induction voltage of the pulsed magnetic fields), was operated at $\approx$ 82 MHz.
    We also performed complementary measurements of the  electrical resistance of \fesete\, under low magnetic fields up to 7\,T in a physical property measurement system (PPMS, Quantum Design).
	Magnetoresistance measurements of \feses\, was performed at High Field Magnet Laboratory (HFML) in Nijmegen.
    For the electrical resistance measurements, we used the conventional four-terminal method with current applied within the $ab$ plane.
    In all the measurements, the magnetic field was applied along the $c$-axis.
    
    \begin{figure}[t]
        \centering
        \includegraphics[width=\linewidth]{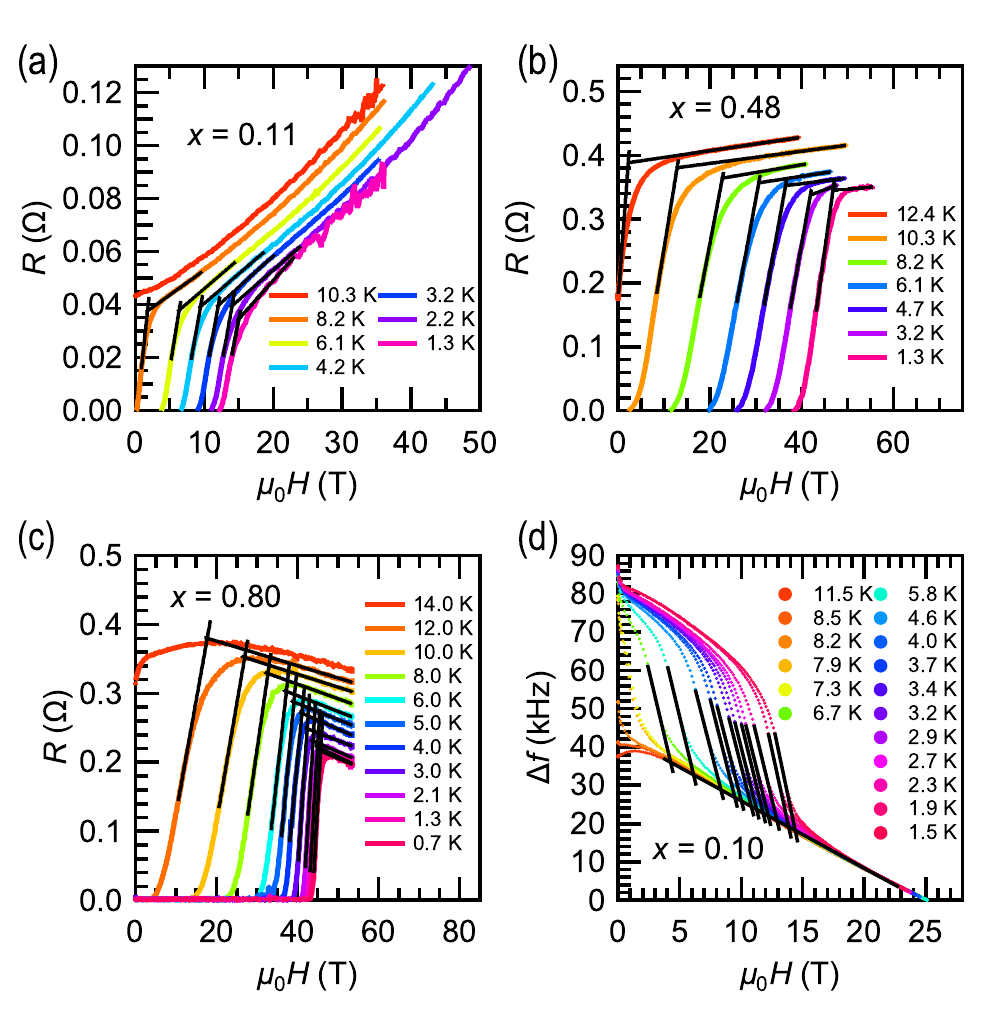}
        \caption{\label{fig:MR_TDO}
		Determination of \hc\, in \fesete.
		\FigCap{a-c} Magnetoresistance measured at several temperatures for $x = 0.11$ \FigCap{a}, $0.48$ \FigCap{b}, and $0.80$ \FigCap{c}.
		The upper critical field \hc\, is defined as an intersection of the two black lines. 
		\FigCap{d} Magnetic field dependence of the change in the resonant frequency $\Delta f$ observed in the TDO measurement for $x = 0.10$.
		\hc\, is determined from the point where two black lines are crossing.
        }
    \end{figure}

	\begin{figure}[t]
        \centering
        \includegraphics[width=\linewidth]{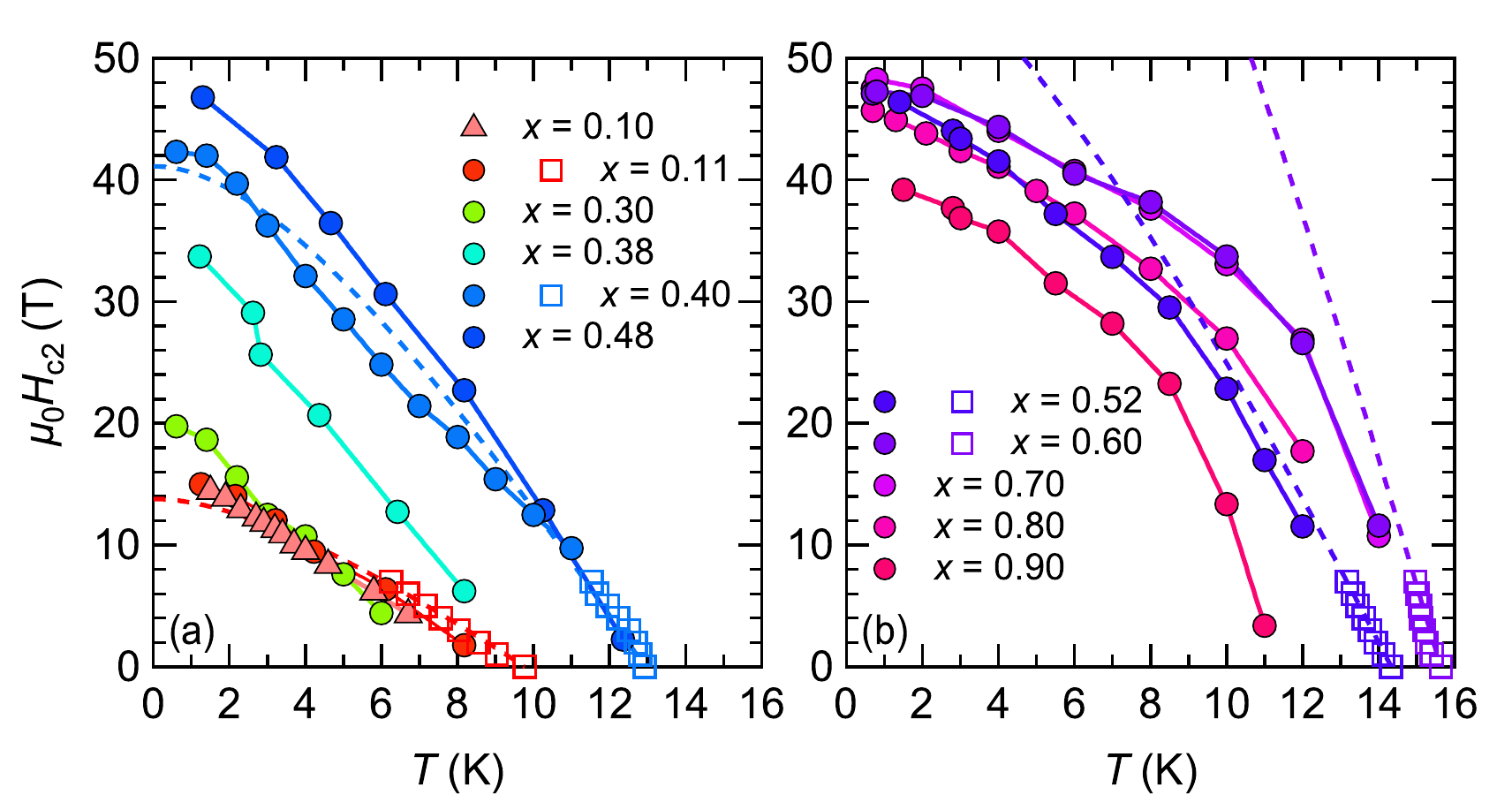}
        \caption{\label{fig:Hc2}
		Temperature dependence of the upper critical field \hc\, in \fesete.\FigCap{a,b} \hc$(T)$ for the compositions \FigCap{a} inside and \FigCap{b} outside of nematic phase, determined from the TDO (triangles) and the resistance  (circles) measurements under pulsed high magnetic fields.
		Solid lines represent the linear interpolation between each data.
		Dashed lines show fits to the low-field data (open squares) measured under the steady field up to 7\,T within the Helfand–-Werthamer framework, which only includes the orbital pair breaking effect.
        }
    \end{figure}
               
    Figure\,\ref{fig:MR_TDO} presents the transverse magnetoresistance data for $x = 0.11$ (\FigCap{a}), $0.48$ (\FigCap{b}), and $0.80$ (\FigCap{c}).
	As $x$ increases, the orbital magnetoresistance in the normal state gradually weakens, and its slope changes from positive to negative
	(See Supplemental Information for all the magnetoresistance data).
	In order to determine the upper critical fields \hc\, for all the compositions regardless of the sign of magnetoresistance, we define \hc\, as the field where two lines extrapolated from the resistance curves in the flux-flow regime and normal state intersect (black lines in Fig.\,\ref{fig:MR_TDO}\FigCap{a}-\FigCap{c}).
	Radio-frequency penetration depth measurements using a TDO were performed for four low $x$ compositions, in which we find noticeable orbital magnetoresistance.
	The shift in the resonant frequency of the TDO circuit $\Delta f$ is proportional to the magnetic penetration depth.
	As shown in the field dependence of $\Delta f$ for $x=0.10$ displayed in Fig.\,\ref{fig:MR_TDO}\FigCap{d}, a clear signature of the superconducting transition is observed and \hc\, is determined from an intersection of two extrapolated lines (See Supplemental Information for TDO data in the other three compositions). 
	The obtained temperature dependence of \hc$(T)$ for the representative Te compositions is shown in Fig.\,\ref{fig:Hc2} (See Supplemental Information for all compositions). 
	The \hc$(T)$ data determined from the magnetoresistance for $x = 0.11$ and from TDO measurements for $x = 0.10$ almost coincide with each other (Fig.\,\ref{fig:Hc2}\FigCap{a}), confirming that our estimates of \hc\, are reasonable.
    In a wide composition range between $x = 0.48$ and $0.70$, the \hc\,values at the lowest temperature are essentially unchanged at 46 - 48\,T, in good agreement with the previous the reports for $x = 0.52$ \cite{Braithwaite2010} and $x = 0.60$ \cite{Khim2010}.
    


	\begin{figure}[t]
		 \centering
		 \includegraphics[width=\linewidth]{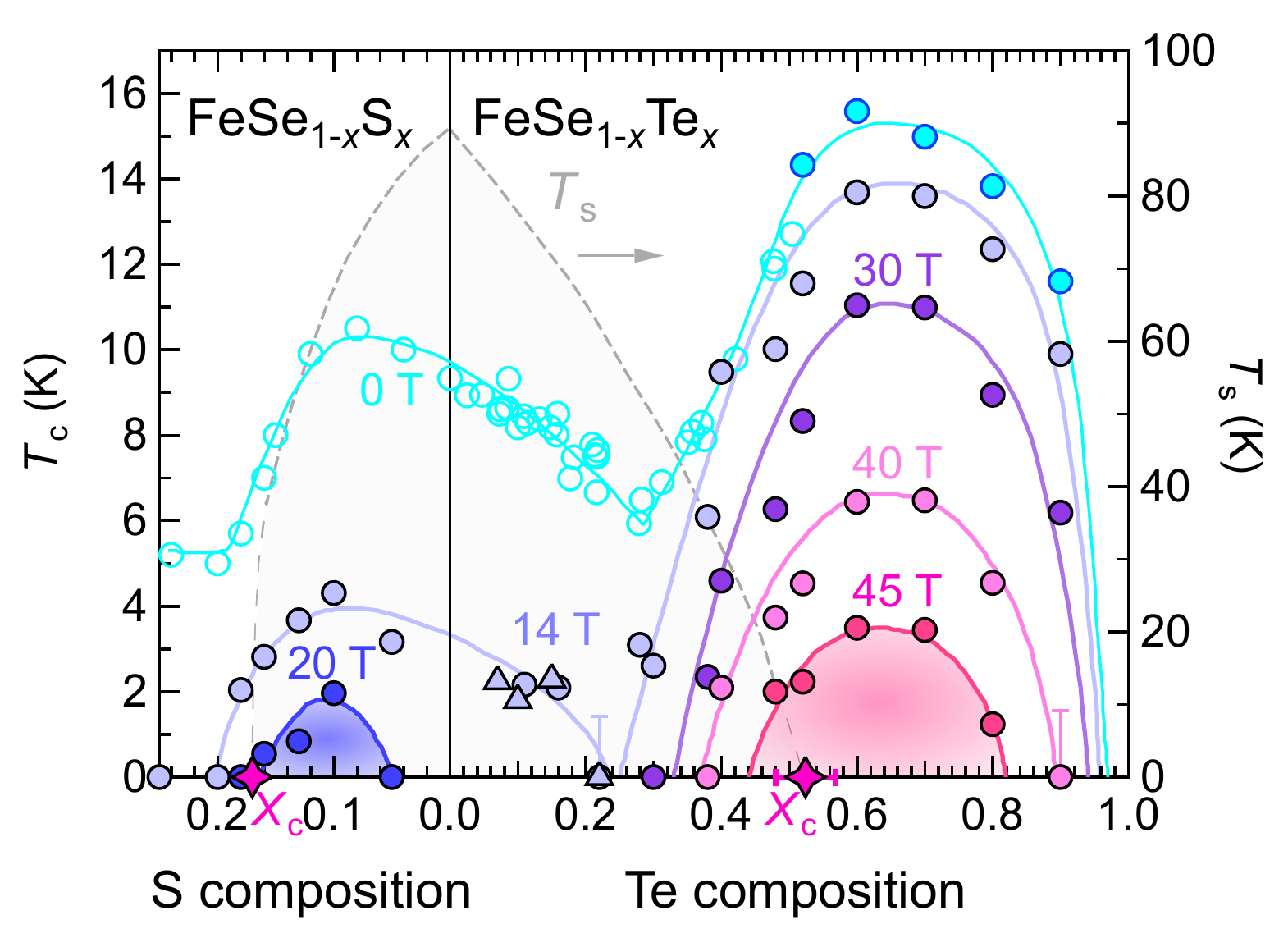}
		 \caption{\label{fig:Tc}
		 Field-induced change of superconducting phases in FeSe-based superconductors.
		 \tc\, values at finite fields are determined from the temperature dependence of \hc\, measured by magnetoresistance (closed circles) and TDO technique (triangles).
		 \tc\, values at zero-field are taken from the data previously measured in the single crystals grown by the CVT technique \cite{Mukasa2021}.
		 Zero-temperature intercepts of the structural transition \ts\, shown by the gray dashed line represent the nematic quantum critical points $x_{\rm c}$.
		 All the solid lines are guides to the eye.
		 }
	 \end{figure}

	From the linear interpolation of the \hc$(T)$ data, we have obtained the \tc\, values at several magnetic field strengths $H$.
    In Fig.\,\ref{fig:Tc}, we show how the magnetic field changes the superconducting phases of \fesete.
	Remarkably, at $\mu_{0}H = 14$\,T, \tc\, for $x < 0.30$ is strongly suppressed, as a result of which the superconducting phase splits into two separated regions.
	Combining the \tc\, values of \feses, obtained by applying the same analysis protocol to its magnetoresistance data \cite{Licciardello2019b,Culo2022}, we find that at $\mu_{0}H = 14$\,T, FeSe exhibits two distinct superconducting domes as a function of the chemical pressure applied by isovalent substitution.
    At $\mu_{0}H = 30$\,T, the superconductivity completely disappears at $x < 0.30$ of FeSe$_{1-x}$Te$_x$, leaving a single superconducting dome centered around $x = 0.60$.
	Moreover, as the magnetic field further increases to $\mu_{0}H = 40$\,T and 45\,T, this dome shrinks to a narrower $x$ region straddling the nematic QCP ($x \approx 0.50$). This demonstrates an intimate relationship between the nematic QCP and the superconducting dome in \fesete. 
	These are the main findings of the present paper.

	The observation of two disconnected superconducting domes, which shrink in a different manner in the presence of a magnetic field, implies that there are two distinct sources of electron pairing in this FeSe-based system. The positions of the dome centers provide valuable clues as to their origin.
	In the lower dome peaked around $x \approx 0.10$ in FeSe$_{1-x}$S$_x$, magnetic fluctuations are likely to play a dominant role since the nematic QCP of \feses\, is pushed outside the dome at 20\,T while several probes have detected substantial stripe-type antiferromagnetic spin fluctuations inside the nematic phase \cite{Wang2016,Wiecki2018}.
	In \fesete\, by contrast, the spin-lattice relaxation rate $1/T_{1}T$ obtained from NMR is shown to be temperature independent for $x = 0.58$ indicating the absence of strong spin fluctuations near the higher dome \cite{Arcon2010}.
	Although the detailed Te composition dependence of spin fluctuations is currently lacking, the pressure-induced magnetic order found in the region including the lower \tc\, dome has been found to disappear below 8\,GPa for $x \geq 0.14$ of FeSe$_{1-x}$Te$_x$ \cite{Mukasa2021}, suggesting that magnetic interactions become weaker in the higher $T_{\rm{c}}$ dome region. 
	Furthermore, nematic fluctuations are expected to be insensitive in this range of magnetic field, as reported by recent elastoresistivity measurements on the electron-doped iron pnictides showing that the nematic susceptibility does not have significant field dependence up to 65\,T \cite{Straquadine2019}.
	Therefore, our observation strongly suggests that nematic fluctuations are mainly responsible for the field-robust superconducting dome centered near the QCP in \fesete.
	Of course, one may contemplate an alternative scenario in which there is a single pairing mechanism underlying each dome with \tc\, depending solely on the density of states participating in the formation of Cooper pairs.
	However, such a scenario would have to be reconciled with the contrasting strength of nematic and magnetic fluctuations across the two domes.

	\begin{figure}[t]
		 \centering
		 \includegraphics[width=0.9\linewidth]{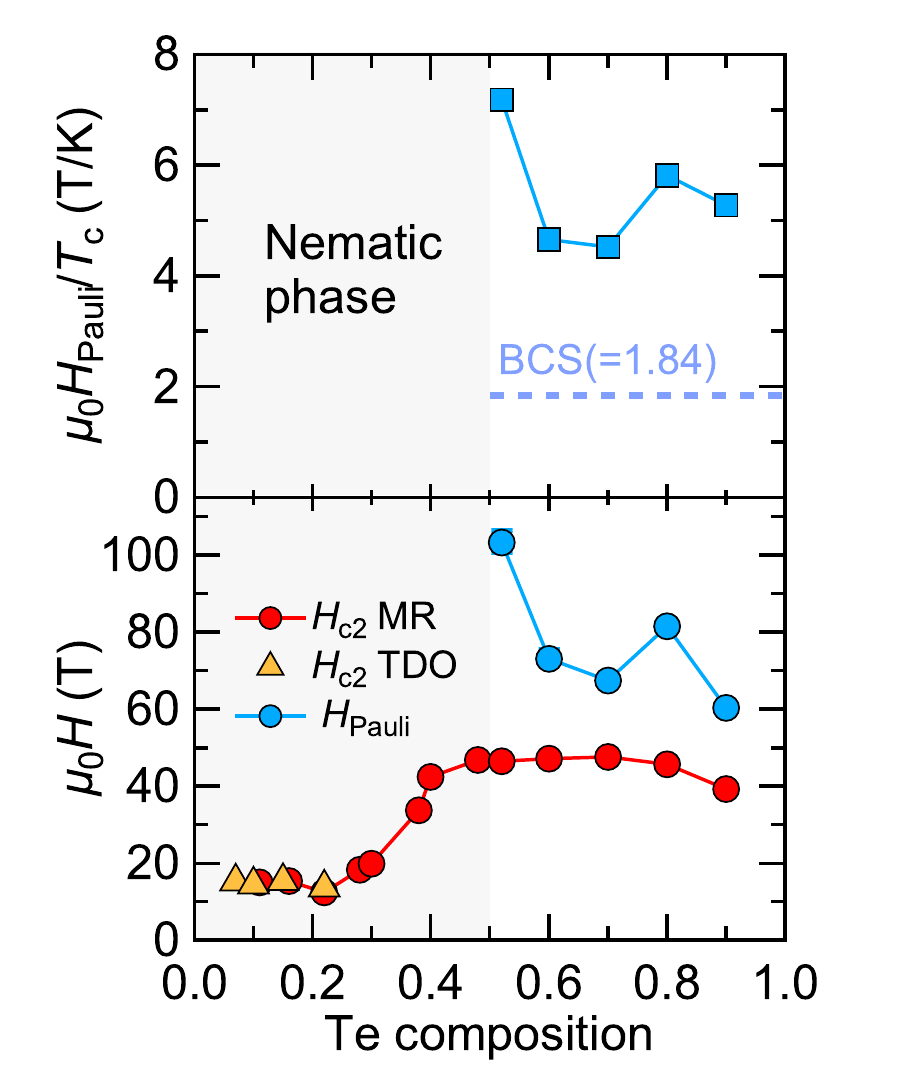}
		 \caption{\label{fig:Hc2/Tc}
		 Te composition $x$ dependence of the upper critical field in \fesete.
		 The bottom panel displays \hc\, at the lowest temperature determined by the resistance (red circles) and TDO (yellow triangles) measurements, and the Pauli-limited field $H_{\rm Pauli}$ (blue circles) estimated as mentioned in the main text. 
		 The top panel shows $H_{\rm Pauli}$ (blue squares) divided by their critical temperature \tc.
		 The dashed line indicates the value predicted in weak-coupling BCS theory.
		 The gray shades represent the compositions where nematic phase appears.}
	 \end{figure}


	More detailed analysis of the upper critical fields can provide us with additional insights into the close relationship between the nematic QCP and pairing strength in \fesete.
	It has been argued that \hc$(T)$ of FeSe and \fesete\, with $x \approx 0.50 - 0.70$ shows Pauli-limited behavior at low temperatures \cite{Braithwaite2010,Khim2010,Her2015}.
	In Fig.\,\ref{fig:Hc2}, we compare the measured values with the curves extrapolated from the low field data (open squares) using the Helfand--Werthamer (HW) formalism, which can yield \hc$(T)$ assuming only the effect of orbital depairing \cite{Helfand1966}.
	For $x<0.40$, there is no strong downward deviation from the HW fitting curves at low temperatures.
	One possible reason for this unexpected behavior is that the resistivity measurements tend to underestimate the initial slope of \hc\, with respect to that thermodynamically determined from, e.g. heat capacity measurements \cite{Coldea2013}.
	In contrast, for $x>0.50$ outside the nematic phase, HW curves clearly go beyond the measured values, indicating a significant paramagnetic pair-breaking effect.
	This is consistent with the nearly isotropic \hc\, at $x=0.67$ \cite{Her2015} whereas the orbital depairing effect should reflect the anisotropic quasi-two dimensional electronic structure of \fesete.
	In this composition range, where the orbital critical field $H_{\rm orb}(0)$ exceeds \hc(0), the Pauli-limited field $H_{\rm Pauli}(0)$ can be estimated using the expression \cite{Khim2011},   
	\begin{equation}
		H_{\rm Pauli}(0) = \frac{\sqrt{2}H_{\rm orb}(0)}{\sqrt{[H_{\rm orb}(0)/H_{\rm c2}(0)]^{2}-1}}.
	\end{equation}
	in which the contribution from spin-orbit coupling is neglected. Using the measured \hc(0) values and the orbital critical field derived from $H_{\rm orb}(0) = -0.69T_{\rm c}\times {\rm{d}}H_{\rm{c}2}/{\rm{d}}T|_{T_{\rm c}}$ (See Supplemental Information), we plot the $x$ dependence of the obtained Pauli-limited field $H_{\rm Pauli}$ in Fig.\,\ref{fig:Hc2/Tc}.
	The value of $H_{\rm Pauli}(0)$ is known to be connected to the magnitude of the zero-temperature superconducting gap $\Delta_{0}$ as $\sqrt{2}\Delta_{0}=g\mu_{\rm B}\mu_{0}H_{\rm Pauli}(0)$, where $g$ and $\mu_{\rm B}$ are the Land\'e $g$ factor and the Bohr magneton, respectively.
	For all compositions, $\mu_{0}H_{\rm Pauli}/T_{\rm{c}}\propto\Delta_{0}/T_{\rm{c}}$, a measure of the pairing strength, is well above $\mu_{0}H_{\rm Pauli}(0)/T_{\rm{c}}=1.84 \,{\rm T/K}$, the value expected for a weak-coupling BCS superconductor, as shown in Fig.\,\ref{fig:Hc2/Tc}. 
	This demonstrates that superconductivity in \fesete\, lies in the strong coupling regime, as discussed previously in the context of the BCS-BEC crossover \cite{Rinott2017}.
	More importantly, we find that $\mu_{0}H_{\rm Pauli}/T_{\rm{c}}$ becomes largest at $x=0.52$.
	Although this analysis cannot be applied to data inside the nematic phase, it clearly indicates that the electron pairing is strengthened upon approach to the nematic QCP in the tetragonal phase.
	It should be noted that there is an increasing trend of the density of states toward FeTe associated with the orbital-selective mass renormalization featuring the strongest correlations in the $d_{xy}$-dominated bands \cite{Yin2011}. 
	This is likely responsible for the observation that the broad \tc$(x)$ dome near the nematic QCP is widened to the high substitution side in \fesete. 
	Moreover, our analysis of the Pauli-limited field strongly suggests that the enhanced pairing interactions near the QCP can cause some pair breaking effect, which results in a slight suppression of \tc\, while maintaining a large $\Delta_{0}$.  

	It is expected that the bosonic interaction strengthening Cooper pairing, which likely corresponds to nematic fluctuations here, can also lead to a strong renormalization of the quasiparticle effective mass $m^{*}$.
	In BaFe$_2$(As,P)$_2$, several probes indicate a diverging $m^*$ towards the antiferromagnetic QCP, where critical spin fluctuations promote \tc\, \cite{Shibauchi2014}.
	However, in nonmagnetic \fesete\, an earlier ARPES study did not detect signatures of mass enhancement in the vicinity of the nematic QCP in any of the orbitals located around the $\Gamma$ point \cite{Liu2015}, consistent with the lack of an enhancement in $H_{\rm orb}(0)$ toward the QCP (see Supplemental Information).
	This absence of any strong enhancement in the effective mass can be attributed to the nemato-elastic coupling effect, which restricts the divergence of the correlation length at the nematic QCP only along high symmetry directions \cite{Paul2017}.
	In FeSe$_{1-x}$S$_x$, in which the nemato-elastic coupling has been estimated to be close to that in \fesete\, \cite{Ishida2022}, a quantum oscillation study found no evidence of mass enhancement near the nematic QCP \cite{Coldea2019}, although non-trivial departures from Fermi liquid  behavior in the magnetotransport have been observed near the QCP \cite{Licciardello2019, Huang2020}.
	An enhanced pairing strength in the vicinity of the nematic QCP in \fesete\, without accompanying apparent mass divergence suggests that nematic and antiferromagnetic fluctuations promote Cooper pairing by a fundamentally different mechanism.
	Together with the field-robust superconducting dome around the nematic QCP, our results imply that \fesete\, could be an ideal playground to study in depth the underlying physics of superconductivity enhanced by nematic critical fluctuations.

	Isovalent substitution is recognized as a relatively clean tuning parameter in iron-based superconductors due to the fact that it replaces elements outside of the Fe layers.
	One intriguing aspect of our observation is that although the electronic structure in the isovalently substituted FeSe maintains electron-hole compensation without introducing strong disorder, the phase diagram displays two distinct superconducting phases.  
	Similar two \tc\, domes have been also found in the phase diagrams of high-\tc\, cuprates \cite{Grissonnanche2014,Ramshaw2015,Chan2020} and heavy fermion compound CeCu$_{2}$Si$_{2}$ \cite{Yuan2003}.
	In these systems just as in the FeSe-based system considered here, one dome is connected to antiferromagnetic fluctuations while the other likely is associated with a QCP of non-magnetic origin, -- the enigmatic pseudogap or charge order in cuprates, the valence transition in CeCu$_{2}$Si$_{2}$, and here, the electronic nematic order in \fesete.
	This might be highlighting the delicate balance of the interactions between spin and other degrees of freedom in these correlated electron systems, and the dominant fluctuations finally lead to the formation of Cooper pairing.

\begin{acknowledgments}
	We thank Y. Mizukami, H. Kontani and I. Paul for the fruitful discussion.
    This work was partially carried out by the joint research in the Institute for Solid State Physics, University of Tokyo. 
	This work was supported by Grants-in-Aid for Scientific Research (KAKENHI Grant Nos JP19H00649 and 18H05227 ), Grant-in-Aid for Scientific Research on Innovative Areas “Quantum Liquid Crystals” (KAKENHI Grant No. JP19H05824) from Japan Society for the Promotion of Science, and by JST CREST (JPMJCR19T5). N. E. H. and M. C. acknowledge support from the Netherlands Organisation for Scientific Research (NWO) (Grant No. 16METL01) —“Strange Metals” and the European Research Council (ERC) under the European Union’s Horizon 2020 research and innovation programme (Grant Agreements No. 835279-Catch-22).
	The work at Hirosaki was supported by Hirosaki University Grant for Distinguished Researchers from fiscal year 2017 to 2018.
\end{acknowledgments}

\bibliographystyle{apsrev4-2}
\bibliography{Hc2.bib}

\end{document}